\documentclass[aps,prb,twocolumn,amsmath,amssymb,superscriptaddress,floatfix,showpacs]{revtex4}

\usepackage{graphicx}

\begin{document}
\title{Exotic Mott phases of the extended t--J model
on the checkerboard lattice at commensurate densities}
\author {Didier Poilblanc}
\affiliation{Laboratoire de Physique Th\'eorique, C.N.R.S. \&
Universit\'e de Toulouse, F-31062 Toulouse, France }
\affiliation{Theoretische Physik, ETH Z\"urich, CH-8093 Z\"urich, Switzerland}
\pacs{75.10.-b, 75.10.Jm, 75.40.Mg, 74.20.Mn, 71.10.Fd}
\begin{abstract}
Coulomb repulsion between electrons moving on a frustrated lattice
can give rise, at simple commensurate electronic densities, to
exotic insulating phases of matter. Such a phenomenon is
illustrated using an extended t--J model on a planar pyrochlore
lattice for which the work on the quarter-filled case
[cond-mat/0702367] is complemented and extended to 1/8- and
3/8-fillings. The location of the metal-insulator transition as a
function of the Coulomb repulsion is shown to depend strongly on
the sign of the hopping. Quite generally, the metal-insulator
transition is characterized by lattice symmetry breaking but the
nature of the insulating Mott state is more complex than a simple
Charge Density Wave. Indeed, in the limit of large Coulomb
repulsion, the physics can be described in the framework of
(extended) quantum fully-packed loop or dimer models carrying
extra spin degrees of freedom. Various diagonal and off-diagonal
plaquette correlation functions are computed and the low-energy
spectra are analyzed in details in order to characterize the
nature of the insulating phases. We provide evidence that, as for
an electronic density of n=1/2 (quarter-filling), the system at
$n=1/4$ or $n=3/4$ exhibits also plaquette order by forming a
(lattice rotationally-invariant) Resonating-Singlet-Pair Crystal,
although with a quadrupling of the lattice unit cell (instead of a
doubling for $n=1/2$) and a 4-fold degenerate ground state.
Interestingly, qualitative differences with the bosonic analog
(e.g. known to exhibit columnar order at n=1/4) emphasize the
important role of the spin degrees of freedom in e.g. stabilizing
plaquette phases w.r.t. rotational symmetry-breaking phases.
\end{abstract}
\date{\today}
\maketitle

\subsection{Introduction}

\begin{figure}
  \centerline{\includegraphics*[angle=0,width=0.8\linewidth]{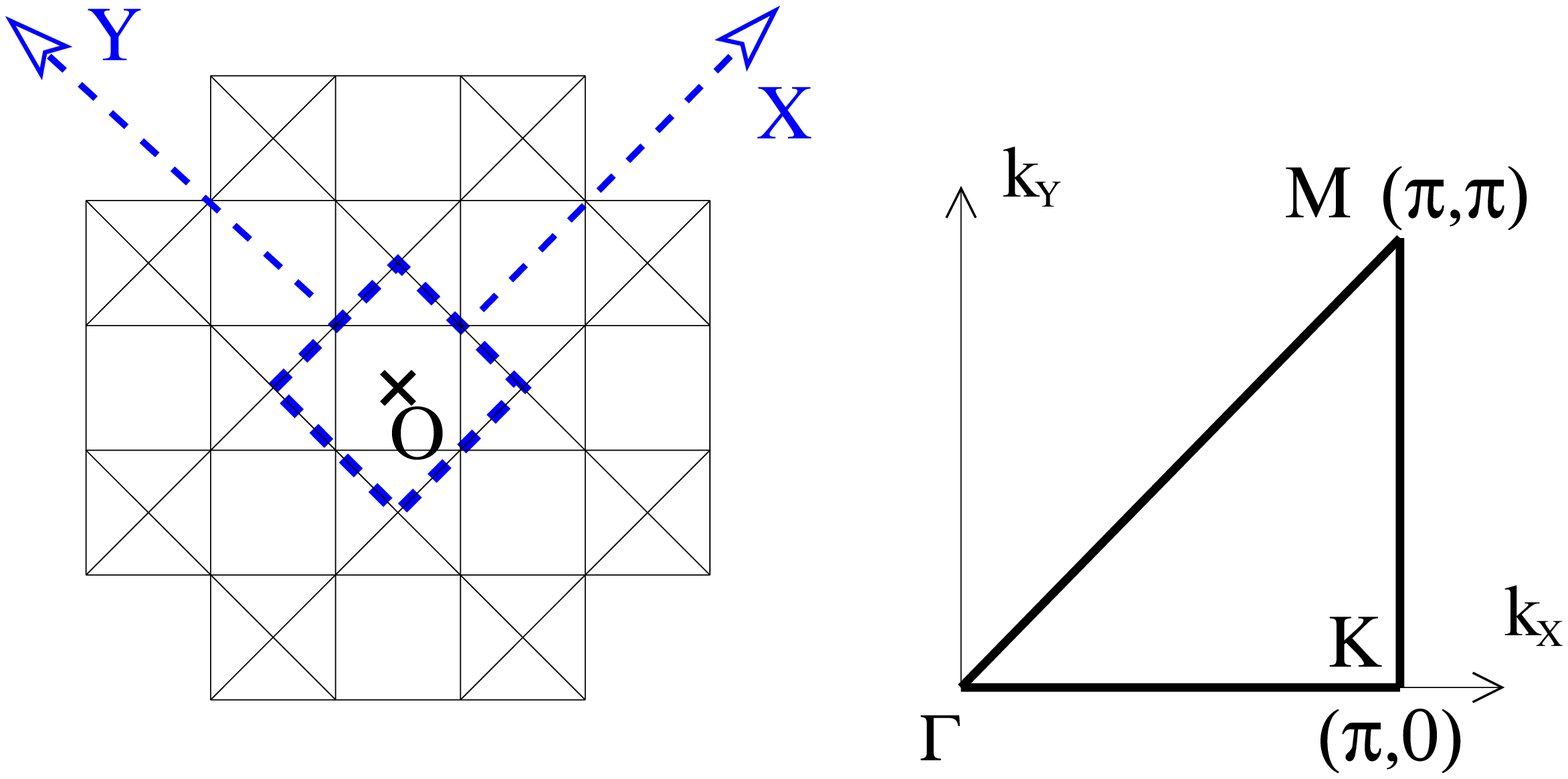}}
  \caption{\label{fig:lattice}
(Color on-line) Left: the pyrochlore lattice is composed of ``crossed plaquettes''
and ``empty squares''. The two-site unit cell
(dashed square)
and the X and Y axes considered in this study are shown. The $C_{4v}$ point group
is defined w.r.t the center O of an empty square. Right: Path
$\Gamma-M-K-\Gamma$ in the first Brillouin zone considered in this study.}
\end{figure}

Frustrating antiferromagnetic (AF) interaction in quantum
magnets~\cite{Review_Misguich} can give rise to various exotic
spin gapped quantum disordered phases which could be realized in a
number of fascinating materials such as Kagome, spinels or
pyrochlore materials. Among these phases, a Valence Bond Crystal
(VBC), the so-called ``plaquette phase'', which spontaneously
breaks lattice symmetry (while preserving rotation symmetry), is
of particular interest and has been identified~\cite{Fouet} in
e.g. the Heisenberg model on the checkerboard lattice, a
two-dimensional (2D) lattice of corner sharing tetrahedra (i.e.
the 2D analog of the three-dimensional pyrochlore lattice as shown
in Fig.~\ref{fig:lattice}).

Sofar, itinerant correlated fermions on frustrated lattices have
been poorly explored and most investigations deal with weakly
doped Mott insulators~\cite{Didier_doped}. A bit more is known for
bosonic systems (which can be mapped on quantum spins under
magnetic field) for which a larger set of numerical techniques is
available (like Quantum Monte Carlo, QMC). On the triangular
lattice, hard-core bosons with nearest-neighbor (NN) repulsive
interaction exhibit a rich phase diagram~\cite{bosons_triangle}
with charge ordering at commensurate 1/3 or 2/3 fillings as well
as a supersolid (i.e. a phase with both superfluid and charge
orders) under light doping. Recently, hard-core bosons with NN
repulsion have also been investigated on the checkerboard lattice
at 1/4-filling where a weakly first-order superfluid-insulator
transition has been detected when increasing Coulomb
repulsion~\cite{Senthil}. Interestingly, in the large repulsion
insulating phase, evidences have been found in favor of lattice
symmetry breaking associated to {\it plaquette ordering} similar
to the one of the Quantum Dimer Model (QDM) on the square
lattice~\cite{RK}. A similar plaquette phase has also been
identified at half-filling in the same model and on the same
lattice using an effective Hamiltonian valid at large
repulsion~\cite{Shannon}. Such a fascinating behavior is in fact
due to the {\it partial} particle localization occurring via
``ice-rule''-like constraints enforced by the minimization of the
global Coulomb energy, a ``classical'' frustrated minimization
problem. It is crucial here that such constraints preserve a
macroscopic degeneracy of the classical configuration manifold.
Indeed, for quarter-filling the Hilbert space is given by the
ensemble of fully packed {\it dimer} coverings (as for the QDM,
see Ref.~\onlinecite{RK}) and a fully packed {\it loop}
representation holds at half-filling~\cite{6VM}. Quantum
fluctuations which occur within these subspaces via second-order
kinetic processes can subsequently lead to exotic
symmetry-breaking states. Here the basic ingredient is the
interplay between the frustrated nature of the lattice and the
Coulomb repulsion (the Ising interaction in the ``spin
language'').

The investigation of the fermionic analog, i.e. fermions moving on
frustrated lattices and subject to (short range) Coulomb repulsion
at commensurate densities, is more involved due the famous
``fermion sign problem'' that prohibits large scale QMC
simulations. However, limited progress have been made recently
thanks to the construction of effective models describing the
low-energy physics at large Coulomb repulsion. These models (or
simple extensions of them) might have known GS in some very
special limits (analogous to the Rokhsar-Kivelson point of
Ref.~\onlinecite{RK}) and, otherwise, are tractable by weak
coupling expansion around a periodic pattern of appropriately
chosen clusters~\cite{Karlo} and by Exact Diagonalisation (ED)
techniques. In the case of spinless fermions on the checkerboard
lattice, the dynamics within the constrained Hilbert space is
introduced via 6-sites ring processes on hexagons involving 3
particles simultaneously and gives rise to a rich phase diagram
~\cite{Pollmann} and, under some conditions, to particle
fractionalization~\cite{fractional}. The case of spinful fermions
which has been  recently investigated at and around
quarter-filling~\cite{Supersolid} involves a second-order kinetic
process on the ``empty squares'' (i.e. the void plaquettes) of the
planar pyrochlore lattice. Interestingly, it was shown that the
quarter-filled GS is a new type of plaquette insulating phase, the
Resonant Singlet Pair Crystal (RSPC): qualitatively, within a
given sublattice of the empty squares (the empty squares of the
checkerboard lattice of Fig.~\ref{fig:lattice} being decomposed
into a ``checkerboard'' pattern of two sublattices), electron
singlet pairs formed on the square diagonals resonate between the
two possible configurations. It was shown that such a plaquette
phase is remarkably robust w.r.t. perturbations (like diagonal
empty square repulsion) contrary to the bosonic case discussed
earlier for which a plaquette phase also exists but with a much
more limited extension in parameter-space. In addition, it has
been argued that lattice symmetry breaking would survive under
light doping in the hole-paired region, giving rise to a fermionic
analog of a supersolid~\cite{bosons_triangle}. It should be
noticed that, besides electronic systems on frustrated lattices
with large Coulomb interactions, cold atoms in optical lattices
offer also a fantastic playground for such ideas since models of
the form discussed in this paper can in  principle be realized by
the appropriate tuning of atomic interactions~\cite{Buechler}.

In this paper, we largely extend the investigation of correlated
fermions on the checkerboard lattice initiated in
Ref.~\onlinecite{Supersolid} by considering, besides $n=1/2$
(quarter-filling), other electron fillings like $n=1/4$
(1/8-filling) and $n=3/4$ (3/8-filling) potentially interesting to
stabilize new insulating phases. We study the metal-insulator
transition within the extended t-J model for increasing Coulomb
repulsion and we introduce then (following
Ref.~\onlinecite{Supersolid}) an effective model describing the
insulating state of matter. In order to characterize the various
insulating phases, (i) we analyze the structure of the low-energy
spectrum in the light of group theory predictions for
symmetry-breaking phenomena and (ii) we compute, for the first
time in the case of itinerant systems, diagonal and off-diagonal
plaquette correlation functions.

Our starting point is the (fermionic) extended Hubbard-like model including
nearest neighbor repulsion $V$ on the checkerboard lattice.
Here, we consider the limit where the on--site repulsion $U$ is
very large (and the largest energy scale), thus forbidding double occupancy via
the Gutzwiller projector ${\mathcal P}_G$
and leading to a t-J-V model description:
\begin{eqnarray}
\mathcal{H}&=&
  -t\sum_{i,j}
  {\mathcal P}_G\left(
     c_{i\sigma}^\dagger c_{j \sigma}^{\phantom{\dagger}}+h.c.
   \right) {\mathcal P}_G + V \sum_{i,j} n_{i}n_{j} +H_J\, ,
\nonumber \\
&H_J&=-J \sum_{\big< ij\big>}\left(\frac{1}{4}-  {\bf S}_{i} \cdot
{\bf S}_{j} \right) n_i n_j \, ,
\label{eq:tJV}
\end{eqnarray}
where $J$ is the AF exchange constant. Hereafter, when dealing
with this model and if not specified, $t=1$ sets the energy scale.
Exact Diagonalisations (ED) have been performed on a periodic
$\sqrt{32}\times\sqrt{32}$ cluster ($N=32$ sites) using full
translation, point group (see Fig.~\ref{fig:lattice}) and
time-reversal ($S_i^Z\rightarrow -S_i^Z$) symmetries~\cite{note1}.

\subsection{Metal-Insulator transition in the t-J-V model}

In the following, we focus on the special electron densities,
$n=k\frac{1}{4}$, where the integer $k$ is 1, 2 or 3, for which we
expect an insulator for sufficiently large ratio $V/|t|$. The
metal-insulator (MI) transition (for increasing $V$) we are
interested in is not directly connected to any nesting properties
of the Fermi surface but it is rather induced by the Coulomb
potential itself (hence it appears for both signs of $t$). To
understand it qualitatively, let us first assume that the NN
repulsion $V$ is the largest energy scale in Hamiltonian
(\ref{eq:tJV}). In that case, the Coulomb energy is minimized by
fulfilling the ``ice rule'' constraint of exactly $k$ electrons in
each tetrahedra. This condition still preserves an extensive (e.g.
macroscopic) manifold of
states~\cite{Shannon,Pollmann,fractional,Supersolid}. Although
quantum fluctuations occur between such configurations (in second
order in $t$), the GS is an insulator since no charge can be
driven between two distant locations in the system without
violating the local constraints. Because of the huge (i.e.
macroscopic) degeneracy of the classical problem ($V=\infty$), it
is clear that the GS should be different from a simple CDW where
particles are localized at fixed locations, and should exhibit a
more subtle and richer nature as seen later on.

\begin{figure}
  \centerline{\includegraphics*[angle=0,width=0.9\linewidth]{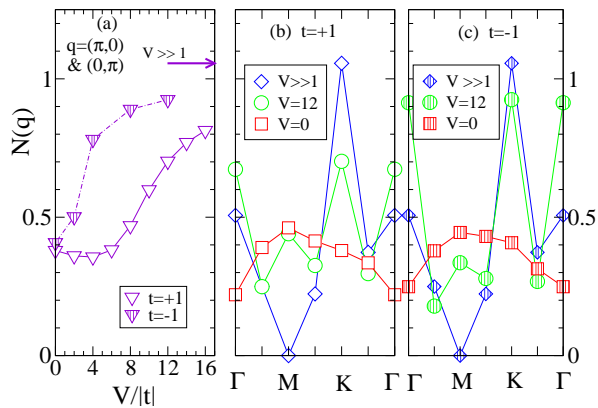}}
  \caption{\label{fig:MIvsV}
(Color on-line) Density structure factor in the GS of the t-J-V
model for $n=1/4$ and $J=0.2$ calculated by ED on a 32-site
cluster. (a) Structure factor at momentum ${\bf  q}=(\pi,0)$ or
$(0,\pi)$ vs $V/|t|$; (b,c) Structure factors as a function of
momentum following the path in the BZ shown in
Fig.~\protect\ref{fig:lattice}. Open (dashed) symbols correspond
to positive (negative) $t$. The points labelled as ``$V>>1$'' are
obtained with the effective model with $W=0$.}
\end{figure}

We have investigated numerically the MI transition as a function of $V$
by computing the charge structure factor
$N({\bf q})=\sum_{\bf R}\exp{(i{\bf q\cdot R})}N(|{\bf R}|)$, the Fourier
transform of the (equal-time) density-density spatial correlation at
distance $R=|{\bf R}|$,
\begin{equation}
N(R)=\frac{1}{N/2} \sum^\prime_{i,j}\{
\big< n_i n_j\big> - n^2 \}   \, ,
\end{equation}
where the sum (over $N/2$ terms, $N$ being the number of sites) is
restricted to sites $i$ and $j$ such that $|{\bf r}_i-{\bf
r}_j|=R$ and ${\bf R}$ has {\it integer} coordinates in terms of
the {\it unit cell lattice vectors} of Fig.~\ref{fig:lattice}
(hence $i$ and $j$ belong to the same sublattice). Note that the
disconnected part has been subtracted for convenience. As shown in
Fig.~\ref{fig:MIvsV} for $n=1/4$ and $J/t=0.2$, characteristic
``Bragg'' peaks at ${\bf q}=(\pi,0)$ and $(0,\pi)$ appear for
increasing $V$ signaling some form of charge ordering
characteristic of a MI transition. However, the complexity of the
insulating phase will be revealed by the simultaneous emergence of
other types of correlations, such as {\it plaquette} correlations
to be discussed later on, suggesting a more subtle and exotic GS
than a simple CDW. We also note that the occurrence of a MI
transition is of course independent on the sign of the hopping
$t$. However, our data show that the rise of the Bragg peak is
more abrupt for negative $t$ and we cannot exclude the MI
transition to be located at $V=0$ in that case (which might be
related to the existence of a flat band at the bottom of the
tight-binding band). However the finiteness of the cluster does
not enable to address the issue of the order of these transitions.

Similar behaviors are of course expected for $n=1/2$ and $n=3/4$
although a direct calculation with Hamiltonian~(\ref{eq:tJV}) is
beyond the power of available computers. Note however that for
$t>0$ perfect nesting of the non-interaction Fermi surface occurs
at $n=1/2$; hence an instability towards a VBC (with
non-equivalent crossed plaquettes) occurs for arbitrary small
interaction~\cite{Indergand} while a different insulating phase is
stabilized at larger $V$ as discussed already in
Ref.~\onlinecite{Supersolid} and in the following sections. An
Insulator-to-Insulator transition is then expected in contrast to
$n=1/4$ and $n=3/4$ (for both signs of $t$) and in contrast to
$n=1/2$ with $t<0$. In all the latter cases, a single MI
transition occurs as depicted above for $n=1/4$, a density for
which a numerical calculations was feasable.

\subsection{Effective model and mapping onto Quantum Dimer or Loop models}

The ice rule constraint (in the insulator at large-V) leads to
simple interesting connections to quantum fully-packed loop or
dimer models~\cite{RK}. Indeed, by associating to each electron a
{\it dimer} joining the centers of the two corner-sharing
tetrahedra it belongs to, a configuration at $n=1/4$ or $n=3/4$
($n=1/2$) can be represented as a dimer (loop) configuration. In
contrast to bosonic analogs~\cite{Shannon}, each dimer now carries
a spin-1/2 degrees of freedom. To fully characterize the
insulating phase, following Ref.~\onlinecite{Supersolid}, we
consider its effective Hamiltonian acting within the constrained
Hilbert space, ${\tilde{\mathcal H}}=H_\square+H_J$ with~:
\begin{eqnarray}
\label{eq:t2}
H_\square&=&-t_2
 \sum_{s} P_\square (s)\, ,
\\
 P_\square (s)&=&\left(
     c_{i\uparrow}^\dagger c_{j\downarrow}^\dagger
    -c_{i\downarrow}^\dagger c_{j\uparrow}^\dagger
  \right)
  \left(
     c_{k\downarrow}^{\phantom{\dagger}}
     c_{l\uparrow}^{\phantom{\dagger}}
    -c_{k\uparrow}^{\phantom{\dagger}}
     c_{l\downarrow}^{\phantom{\dagger}}
  \right) \\
  &+& \left(
     c_{k\uparrow}^\dagger c_{l\downarrow}^\dagger
    -c_{k\downarrow}^\dagger c_{l\uparrow}^\dagger
  \right)
  \left(
     c_{i\downarrow}^{\phantom{\dagger}}
     c_{j\uparrow}^{\phantom{\dagger}}
    -c_{i\uparrow}^{\phantom{\dagger}}
     c_{j\downarrow}^{\phantom{\dagger}}
  \right)\,
  \nonumber
\end{eqnarray}
where $t_2=\frac{2t^2}{V}$ and the summation is over the void
plaquettes labelled by $s$ of the checkerboard lattice. The sites
of $s$ are ordered as $i$,$k$,$j$ and $l$ in the clockwise (or
anti-clockwise) direction. $P_\square (s)$ is a second-order
process which preserves the ice rule. It acts on two electrons
forming a singlet bond on one of the two diagonals of $s$, and
rotates the bond by 90$^\circ$ degree.

\subsection{Description of the expected ordered phases}

\begin{figure}
  \centerline{\includegraphics*[angle=0,width=0.9\linewidth]{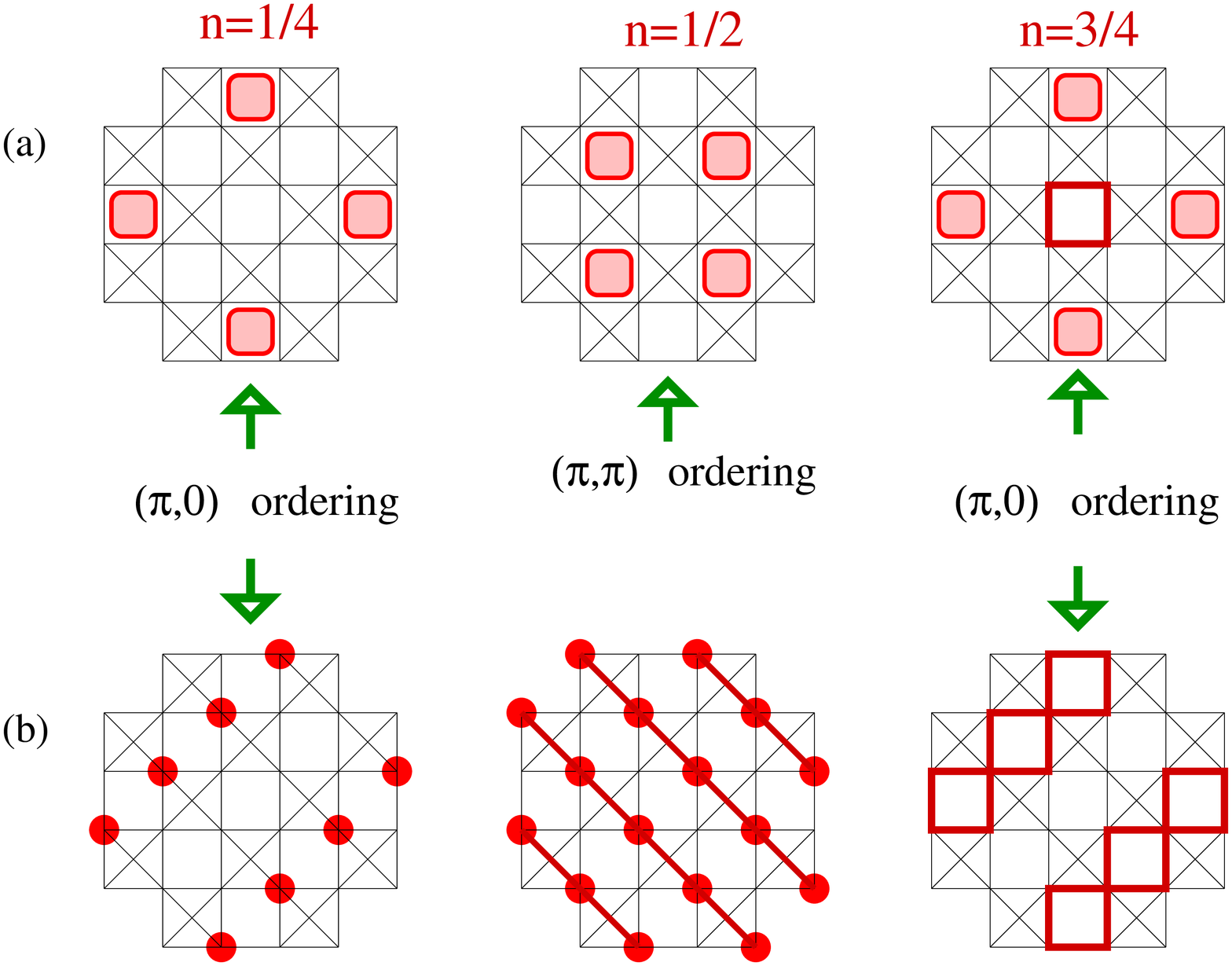}}
  \caption{\label{fig:phases}
(Color on-line) Schematic views of the plaquette (a) and columnar
(b) phases considered in this work for electron
densities $n=1/4$, $1/2$ and $3/4$ (as shown).
Dots, shaded plaquettes
and (red) thick lines correspond to electrons, singlet pairs resonating
on a plaquette and AF bonds, respectively.}
\end{figure}

The kinetic processes of Eq.~(\ref{eq:t2}) favor singlet electron
pairs resonating on void plaquettes (empty squares) and could lead
to long-range plaquette-plaquette correlations.
Fig.~\ref{fig:phases}(a) shows pictorially the expected Resonant
Singlet Pair Crystals exhibiting such a plaquette ordering. For
$n=1/2$ it was indeed shown that the ${\bf q}=(\pi,\pi)$ RSPC is
stable~\cite{Supersolid} as long as $J/t_2< 1.5$. Note that the GS
of the RSPC is two-fold degenerate at $n=1/2$ while one expects a
four-fold degenerate GS at $n=1/4$ and $n=3/4$ in connection with
${\bf q}=(\pi,\pi)$, $(\pi,0)$ and $(0,\pi)$ plaquette order as
seen later. In addition, the average charge density of the $n=1/2$
RSPC is uniform while ${\bf q}=(\pi,0)$ and $(0,\pi)$ (charge
density) Bragg peaks should be present at $n=1/4$ and $n=3/4$
consistently with the charge ordering observed previously at the
MI transition. However, a full characterization of the insulating
phases is delicate since the RSPC of Fig.~\ref{fig:phases}(a) in
fact compete with the columnar-like phases of
Fig.~\ref{fig:phases}(b) which optimize the numbers of possible
t2- (or dimer-) flips. For $n=1/4$ and $n=3/4$ {\it both} RSPC and
columnar phases exhibit $(\pi,0)$ and $(0,\pi)$ plaquette (and
charge) ordering as shown later. However, at $n=1/2$, {\it only}
the RSPC exhibits $(\pi,\pi)$ plaquette ordering. Similarly, only
the columnar states break rotation symmetry (for all commensurate
densities).

The various candidate phases hence correspond to different
symmetry-breaking phenomena which should be directly reflected in
the structure of the low-energy spectra. Table~\ref{table1}
summarizes the quantum numbers of the quasi-degenerate states
which should collapse onto the GS in the thermodynamic limit.
Indeed, from linear combinations of all the equivalent translated
and/or rotated patterns obtained from each of the representative
ones in Fig.~\ref{fig:phases}, one can easily build the
(orthogonal) states carrying the appropriate quantum numbers
reported in Table~\ref{table1}. Analyzing the low-energy energy
spectra should then give clear signals on the nature of the
symmetry-breaking phase.

\begin{table}
\begin{tabular}{|c|c|c|c|c|c|c|}
Ordered Phase& $(0,0)$ & $(0,0)$ & $(\pi,\pi)$ &
$(\pi,0)$ & $(\pi,0)$ & $(\pi,\pi)$ \\
 & $A_1$ & $B_1$ & $A_1$ & $A_1$ & $A_1$ $(*)$ & $B_1$ \\
\hline
$n=1/2$ RSPC & X & & X & & & \\
$n=1/2$ Colum. & X & X & & & & \\
$n=1/4$, $3/4$ RSPC & X & & X & X & & \\
$n=1/4$, $3/4$ Colum.& X & X & & X & & \\
$n=1/4$, $3/4$ Mixed & X & X & X & X & X & X\\
\end{tabular}
\caption{Quantum numbers of the degenerate GS of the various
ordered phases considered here. Note that momentum $(\pi,0)$  is
degenerate with $(0,\pi)$ (not shown). Standard notations are used
to denote the irreducible representations of the $C_{4v}$ ($C_{2v}$)
point group (acting around point O in
Fig.~\protect\ref{fig:lattice})
for momentum $(0,0)$ and $(\pi,\pi)$ ($(\pi,0)$).
(*) stands for a {\it second} $\{(\pi,0),A_1\}$ state.
The RSPC, columnar and mixed GS are 2-, 4- and 8-fold degenerate
respectively.}
\label{table1}
\end{table}

\subsection{Analysis of the low-energy spectra: hints for symmetry-breaking GS}

\begin{figure}
  \centerline{\includegraphics*[angle=0,width=0.9\linewidth]{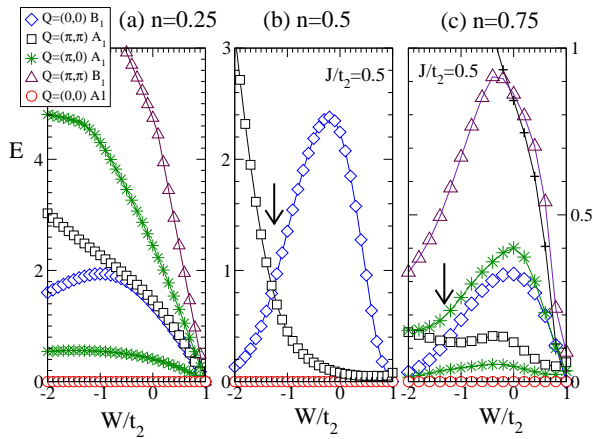}}
  \caption{\label{fig:ener_vsW}
(Color on-line) Lowest excitations (w.r.t. the GS energy) vs $W/t_2$ for
$n=1/4$, (a) $1/2$ (b) and $3/4$ (c) obtained by ED of a 32-site cluster.
Different symbols
correspond to different irreducible representations as shown in
the legend. Other excitations (+ symbols in (c))
have significantly higher energies.
In (b) and (c) the vertical arrows indicate the
level crossings between
the $\{(0,0),B_1\}$ and $\{(\pi,\pi),A_1\}$ states characteristic
of the columnar state and the RSPC respectively. }
\end{figure}

Prior to the detailed analysis of the low-energy spectra it is
useful to carry on the connection to the usual Quantum Dimer Model
(QDM)~\cite{RK} and, following Ref.~\onlinecite{Supersolid}, add
to ${\tilde{\mathcal H}}$ a ``diagonal'' term corresponding to the
4-site ring exchange:
\begin{eqnarray}
\mathcal{H}_W  &=& W \sum_\square
 \big[ \left(\frac{1}{2}-2  {\bf S}_{i} \cdot {\bf S}_{j}
  \right) n_i n_j (1-n_l)(1-n_k) \nonumber \\
  &+& \left(\frac{1}{2}-2  {\bf S}_{k} \cdot {\bf S}_{l}
  \right) n_k n_l (1-n_i)(1-n_j) \big] \,,
\label{eq:W}
\end{eqnarray}
the summation is again over the empty squares $s=(i,k,j,l)$ and
the notations are the same as in Eq.~(\ref{eq:t2}). The
Hamiltonian at the Rohksar-Kivelson point (i.e. for
$W=t_2$)~\cite{RK,Supersolid} is a sum of projectors. Although our
original problem corresponds to $W=0$, analyzing the evolution of
the system from large negative $W$ value (for which a columnar
phase is stabilized) up to the RK point offers new insights to
characterize the $W=0$ phase (we shall not consider here the case
$W>t_2$ where localized states are stabilized). This strategy has
been applied successfully in Ref.~\onlinecite{Supersolid} for
$n=1/2$ and we extend it here to the more complex $n=1/4$ and
$n=3/4$ cases. Fig.~\ref{fig:ener_vsW} shows the low-energy
spectra vs $W/t_2$ for all three commensurate electron densities
considered here. For $n=1/2$ and $n=3/4$, increasing $W$ from
negative value upwards, a level crossing is found around
$W/t_2=w_c\simeq -1.2$ signaling a (possibly first-order) phase
transition. Indeed, our results in Fig.~\ref{fig:ener_vsW}(b,c)
compared to the GS degeneracy of the ordered states listed in
Table~\ref{table1} strongly suggest that the columnar phase (the
RSPC) is stable for $W/t_2<w_c$ ($W/t_2>w_c$) since a group of two
or four quasi-degenerate states (with exactly the expected quantum
numbers) separated from the rest of the spectrum by a small gap
can clearly be identified. In contrast, no level crossing occurs
for $n=1/4$ so that no conclusion can be drawn yet for this case
(the 8-fold degenerate "mixed" state of Table~\ref{table1} not
being completely excluded at this point). For $n=1/2$ and $n=3/4$,
the ``physical points'' $W=0$ are clearly located in the region of
stability of the RSPC.

\subsection{Off-diagonal plaquette-exchange correlation function}

In order to get complementary fingerprints associated to a given
type of symmetry-breaking and more insights on the nature of the
insulating phases, it is of interest to compute the
plaquette-exchange structure factor $C_{\rm pl}({\bf q})=\sum_{\bf
R}\exp{(i{\bf q\cdot R})}C_{\rm pl}(|{\bf R}|)$, the Fourier
transform of the spatial plaquette correlations at distance
$R=|{\bf R}|$,
\begin{eqnarray}
\label{eq:plaquette} C_{\rm
pl}(R)&=&\frac{1}{N/2}\sum^\prime_{s,s'}\{
\big< P_\square(s)P_\square(s')\big> \\
&-& \big< P_\square(s)\big> \big< P_\square(s')\big>\}   \,
,\nonumber
\end{eqnarray}
where the sum is restricted to plaquettes such that $|{\bf
R}_s-{\bf R}_s'|=R$ and we have again subtracted the disconnected
part. Despite the intrinsic complexity of computing a 8-fermion
correlator, the implementation is made easier by the fact that the
operator to average has the full translation and rotation symmetry
of the lattice.

\begin{figure}
  \centerline{\includegraphics*[angle=0,width=0.9\linewidth]{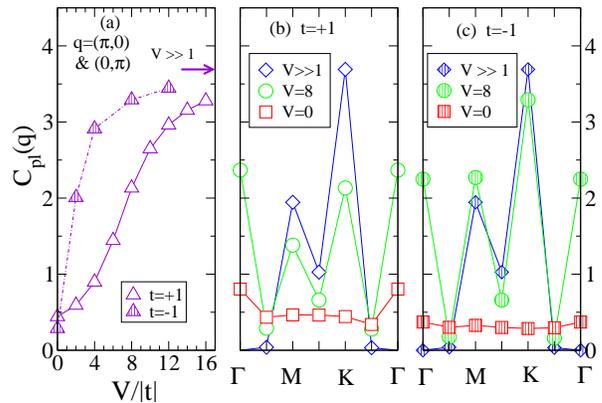}}
  \caption{\label{fig:MIvsV_PL}
(Color on-line) Plaquette-exchange structure factor in the GS of
the t-J-V model for $n=1/4$ and $J=0.2$ obtained by ED of a
32-site cluster. (a) Structure factor at momentum ${\bf
q}=(\pi,0)$ or $(0,\pi)$ vs $V/|t|$; (b,c) Structure factor as a
function of momentum following the path in the BZ shown in
Fig.~\protect\ref{fig:lattice}. Open (dashed) symbols correspond
to positive (negative) $t$. The points labelled as ``$V>>1$'' are
obtained with the effective model with $W=0$.}
\end{figure}

We have first studied the emergence of superstructure peaks in the
plaquette structure factor on the original t-J-V model itself and
the results at 1/8-filling are shown in Fig.~\ref{fig:MIvsV_PL}.
For $V=0$ the correlations have no structure in momentum space.
However, for increasing $V$, superstructure peaks at momenta
$(\pi,0)$ and $(0,\pi)$ (as well as harmonics at $(\pi,\pi)$)
clearly develop. The behavior of the  $(\pi,0)$ peak with $V$
follows quite faithfully the behavior of the ``charge'' Bragg peak
of Fig.~\ref{fig:MIvsV} indicating that charge and plaquette
orderings seem to be tightly connected.

\begin{figure}
  \centerline{\includegraphics*[angle=0,width=0.9\linewidth]{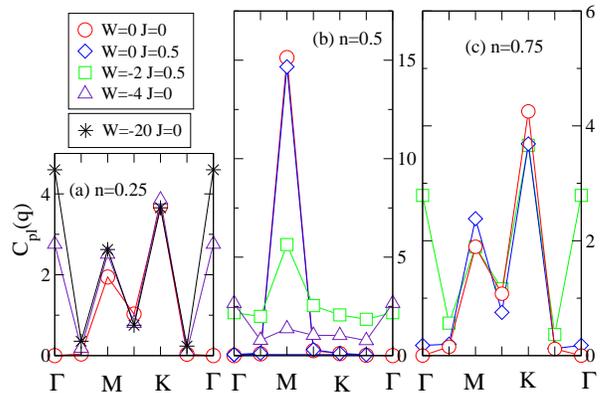}}
  \caption{\label{fig:correl}
(Color on-line) Plaquette-exchange structure factors in the GS of
the effective $t_2-J-W$ model vs momentum ${\bf q}$ along the path
shown in Fig.~\protect\ref{fig:lattice} and for different choices
of $W$ and $J$ as indicated in the legend (in units of $t_2$).
(a), (b) and (c) correspond to commensurate fillings $n=0.25$,
$n=0.5$ and $n=0.75$, respectively. Results are obtained by ED of
a 32-site cluster.}
\end{figure}

Next, we turn to the case of the effective model describing the
insulating phase and results are shown in Fig.~\ref{fig:correl}.
At quarter-filling ($n=1/2$), the large peak emerging at
$(\pi,\pi)$ is directly linked to the existence of the RSPC (at
$W=0$) while it is suppressed in the columnar phase at
sufficiently negative $W$ values, $W/t_2<w_c$, where translation
symmetry is restored. On the contrary, for $n=1/4$ and $3/4$,
since both columnar and RSPC phases break translation symmetry
superstructure peaks at momenta $(\pi,0)$, $(0,\pi)$ and
$(\pi,\pi)$ are always present irrespective of the value of $W$.

\subsection{Diagonal $B_1$-plaquette correlation function: hints for
rotational symmetry breaking}

The analysis of the low-energy spectrum strongly suggests a
transition from the columnar phase to the RSPC by varying $W$, at
least for $n=1/2$ and $n=3/4$. We therefore expect to be able to
detect directly the signal of the columnar order via the emergence
of some long-range correlations. Note that the above investigation
of the plaquette-exchange structure factor, by detecting lattice
translation symmetry breaking, was able to discriminate between
columnar and RSPC phases only for $n=1/2$. Since the rotational
symmetry-breaking in the columnar phase is achieved via the
collapse of a $\{{\bf q=0}, B_1\}$ state onto the GS in the
thermodynamic limit, it is natural to define a diagonal
$B_1$-plaquette correlator as,
\begin{equation}
C_{\rm pl}^{B_1}(R)=\frac{1}{N/2}\sum^\prime_{s,s'}
\big< P_\square^{B_1}(s)P_\square^{B_1}(s')\big> \, ,
\end{equation}
with the diagonal plaquette operator defined as,
\begin{equation}
P_\square^{B_1}(s)=n_i n_j - n_k n_l  \, ,
\end{equation}
and the rest of the notations is the same as in Eqs.~(\ref{eq:t2})
and \ref{eq:plaquette}. Note that there is no need here to
subtract the disconnected part since it is vanishing for obvious
symmetry reasons.

\begin{figure}
  \centerline{\includegraphics*[angle=0,width=0.9\linewidth]{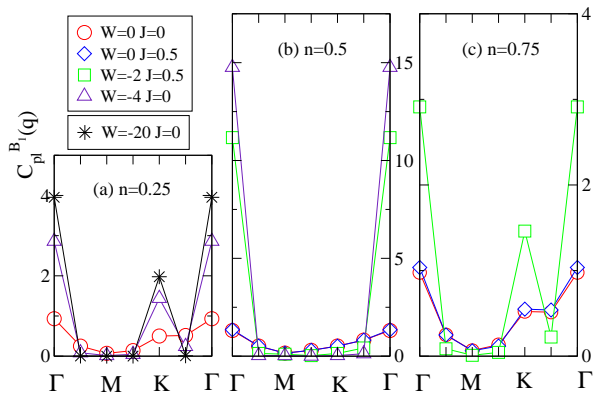}}
  \caption{\label{fig:correl2}
(Color on-line)
$B_1$-plaquette
structure factors in the GS of the effective $t_2-J-W$ model
vs momentum ${\bf q}$
along the path shown in Fig.~\protect\ref{fig:lattice}.
Same parameters and notations as Fig.~\protect\ref{fig:correl}.}
\end{figure}

The results for the related plaquette structure factor (defined as
before) vs momentum are shown in Fig.~\ref{fig:correl2} and
clearly reveal the appearance of a strong ${\bf q=0}$ peak (with
an additional peak at $(\pi,\pi)$ for $n=1/4$ and $n=3/4$) in the
columnar phase i.e. for sufficiently negative $W$ values,
$W/t_2<w_c$. In contrast, for $W=0$, which is the case relevant to
the original t--J--V model, the $B_1$-plaquette structure factor
remains very flat in momentum space consistently with the
existence of a rotationally-symmetric RSPC for all three values of
the densities studied here. This clearly confirms our previous
results for $n=1/2$ and $n=3/4$ and suggests strongly that the
case $n=1/4$ is similar. We therefore attribute the lack of level
crossing in the previous spectral analysis for $n=1/4$ to a finite
size effect.

\subsection{Comparison to the bosonic case}

It is of interest to compare our results to the bosonic system
(again with infinite on-site repulsion and large NN repulsion)
whose effective Hamiltonian is given by (\ref{eq:t2}) where
$P_\square(s)$ is now replaced by $P_\square(s)=b_i^\dagger
b_j^\dagger b_k b_l$ in terms of the usual {\it hard-core} boson
creation and annihilation operators on the empty square
$(i,j,k,l)$. For $n=1/2$ the bosonic system can be mapped on the
usual fully-packed quantum loop model (or a quantized six-vertex
model) with only kinetic processes (i.e. $W=0$) exhibiting a
plaquette phase~\cite{Shannon} bearing some similarities with the
fermionic RSPC~\cite{Supersolid}. However, the stability of the
plaquette phase w.r.t. additional parameters (e.g. $W$) is more
fragile for bosons. The $n=1/4$ and $n=3/4$ cases reduce, for
bosons, to the usual QDM~\cite{RK} studied recently by extensive
QMC methods~\cite{Syljuasen} and believed to have a columnar GS in
the absence of diagonal interactions (i.e. $W=0$). This is
different from our results for spinful fermions exhibiting a
rotationally-invariant RSPC GS. This emphasizes the special role
of the spin degrees of freedom in stabilizing plaquette phases.
Note however that Ref.~\onlinecite{Senthil} reports for the
hard-core bosons system at $n=1/4$, increasing the NN $V$, a MI
transition at $V/t\sim 3.2$ towards a {\it plaquette} phase
different from the supposed GS in the $V/t>>1$ limit described by
the QDM~\cite{Syljuasen}. The evolution from the MI point to
larger $V$ needs certainly to be clarified.

\subsection{Conclusions}

To summarize, using a t-J model extended with NN Coulomb
repulsion, the metal-insulator transition has been investigated on
the planar pyrochlore frustrated lattice at 1/8-filling (i.e.
$n=1/4$), a tractable case numerically. For increasing repulsion,
the emergence of superstructure (charge) Bragg peaks signals the
onset of the insulating phase which also immediately develops some
form of plaquette ordering. To understand further the complex
nature of the insulating phase, an effective model valid for
commensurate fillings like $n=1/4$, $n=1/2$ and $n=3/4$ and in the
limit of large NN repulsion is used. Remarkably, the frustrated
nature of the lattice leads then to ice rule-like constraints so
that the Hilbert space can be mapped onto a manifold of
fully-packed dimer or loop coverings with extra spin degrees of
freedom. Quantum fluctuations take the form of simple
(second-order) kinetic processes that bear strong resemblance with
the usual bosonic QDM. Among the various type of lattice
symmetry-breaking states, several compelling evidences are given
in favor of a plaquette GS, the RSPC. This new state of matter
introduced earlier in the quarter-filled case~\cite{Supersolid},
is extended here to 1/8 and 3/8 fillings for which it has a twice
larger supercell and a twice larger GS degeneracy. A detailed
analysis of the low-energy spectra supplemented by symmetry
considerations as well as the calculation of several types of {\it
plaquette} correlations support our claims. To carry on the
analogy with the QDM, following Ref.~\onlinecite{Supersolid}, we
also introduced a diagonal plaquette interaction and showed
signatures of a (presumably first order) transition towards a
columnar phase as this parameter is tuned to more negative values.

\subsection{Acknowledgements}

I thank the {\it Agence Nationale de la Recherche} (France) for
support, IDRIS (Orsay, France) for computer time and the
Theoretical Physics laboratory at ETH-Z\"urich for warm
hospitality. I am also happy to thank Manfred Sigrist and Matthias
Troyer for valuable discussions and I am indebted to Karlo Penc
and Nic Shannon for patiently explaining basic ideas and
(eventually) getting me interested in this problem.


\begin{thebibliography}{99}
\bibitem{note1} The largest Hilbert space was obtained
for $n=3/4$ and $\{(\pi,0),A_1\}$ symmetry (5,757,416 symmetrized states).
\end{thebibliography}
\end{document}